\documentclass[conference,10pt]{IEEEtran}
\IEEEoverridecommandlockouts

\usepackage{epsfig,rotating,setspace,latexsym,amsmath,epsf,amssymb,amsfonts,bm,theorem,cite,algorithm,graphicx,epsf,authblk,epstopdf,color,algpseudocode,bbm}

\newtheorem{definition}{Definition}
\newtheorem{theorem}{Theorem}

\allowdisplaybreaks

\begin{document}

\title{Online Timely Status Updates with Erasures for Energy Harvesting Sensors\thanks{This research was supported in part by the National Science Foundation under Grants CNS-1702808, ECCS-1650299, CNS-1526608 and ECCS-1807348.}}


\author[1]{Ahmed Arafa}
\author[2]{Jing Yang}
\author[3]{Sennur Ulukus}
\author[1]{H. Vincent Poor}
\affil[1]{\normalsize Electrical Engineering Department, Princeton University}
\affil[2]{\normalsize School of Electrical Engineering and Computer Science, The Pennsylvania State University}
\affil[3]{\normalsize Department of Electrical and Computer Engineering, University of Maryland}

\maketitle

\begin{abstract}
An energy harvesting sensor that is sending status updates to a destination through an {\it erasure} channel is considered, in which transmissions are prone to being erased with some probability $q$, independently from other transmissions. The sensor, however, is {\it unaware} of erasure events due to lack of feedback from the destination. Energy expenditure is normalized in the sense that one transmission consumes one unit of energy. The sensor is equipped with a {\it unit-sized} battery to save its incoming energy, which arrives according to a Poisson process of unit rate. The setting is online, in which energy arrival times are only revealed causally after being harvested, and the goal is to design transmission times such that the long term average age of information (AoI), defined as the time elapsed since the latest update has reached the destination {\it successfully}, is minimized. The optimal status update policy is first shown to have a {\it renewal} structure, in which the time instants at which the destination receives an update successfully constitute a renewal process. Then, for $q\leq\frac{1}{2}$, the optimal renewal policy is shown to have a {\it threshold} structure, in which a new status update is transmitted only if the AoI grows above a certain threshold, that is shown to be a decreasing function of $q$. While for $q>\frac{1}{2}$, the optimal renewal policy is shown to be {\it greedy}, in which a new status update is transmitted whenever energy is available.
\end{abstract}

\section{Introduction}

\begin{figure}
\center
\includegraphics[scale=.9]{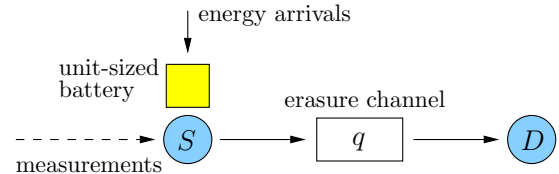}
\caption{Energy harvesting sensor with a unit-sized battery sending measurement status updates through an erasure channel.}
\label{fig_sys_mod}
\vspace{-.25in}
\end{figure}

We consider an energy harvesting sensor monitoring some physical phenomenon and sending measurement status updates to a destination through an erasure channel, see Fig.~\ref{fig_sys_mod}. We design optimal policies that minimize the long term average AoI, with only causal knowledge of energy arrival times and with no erasure error feedback. The AoI is the time elapsed since the latest update has reached the destination successfully.

The AoI metric has been studied in the literature under various settings; mainly through modeling the update system as a queuing system and analyzing the long term average AoI, and through using optimization tools to characterize optimal status update policies, see, e.g., \cite{yates_age_1, yates_age_mac, ephremides_age_random, ephremides_age_non_linear, shroff_age_multi_hop, modiano-age-bc, najm-age-multistream, sun-age-mdp, yates-age-erase-code, yates-age-cache, sun-weiner}, and also the recent survey in \cite{kosta-age-monograph}. AoI minimization in energy harvesting communications has recently gained attention in \cite{yates_age_eh, elif_age_eh, arafa-age-2hop, arafa-age-var-serv, elif-age-Emax, jing-age-online, jing-age-error-infinite-no-fb, jing-age-error-infinite-w-fb, baknina-age-coding, shahab-age-online-rndm, baknina-updt-info, arafa-age-sgl, arafa-age-rbr, arafa-age-online-finite, elif-age-online-threshold} under various service time (time for the update to take effect), battery capacity, and channel assumptions. With the exception of \cite{arafa-age-var-serv}, an underlying assumption in these works is that energy expenditure is normalized, i.e., sending one status update consumes one energy unit. Under a perfect channel model, optimality of online threshold policies, in which a new status update is transmitted only if the AoI grows above a certain threshold, is first shown in \cite{jing-age-online} for unit-sized batteries, and is later extended in \cite{arafa-age-sgl, arafa-age-rbr, arafa-age-online-finite}, and independently and concurrently in \cite{elif-age-online-threshold}, for general finite-sized batteries. Under an erasure channel model, references \cite{jing-age-error-infinite-no-fb, jing-age-error-infinite-w-fb} show the optimality of best-effort uniform updating policies, in which an update is sent every one time unit only if energy is available, for infinite batteries, with and without erasure error feedback. The use of coding to combat erasures along with best-effort and save-and-transmit online policies is studied in \cite{baknina-age-coding}, extending the non-energy-harvesting work in \cite{yates-age-erase-code}. 

In this paper, we extend the results of \cite{jing-age-error-infinite-no-fb} to the case of {\it finite} batteries of unit size. Under Poisson energy arrivals with unit rate, with only causal knowledge of their arrival times, we design age-minimal status update policies for an {\it erasure} communication channel. With the sensor only knowing the erasure probability, and not the actual erasure state per transmission due to {\it lack of feedback} from the destination (receiver), we show that the optimal policy has a {\it renewal} structure. Specifically, the sensor should design its transmission times such that the times at which the destination receives an update {\it successfully} form a renewal process with i.i.d. inter-durations. Then, we show that the optimal renewal policy's structure depends on the value of the erasure probability, $q$. Specifically, if $q\leq\frac{1}{2}$, the optimal policy is a {\it threshold} policy, in which a new update is transmitted only if the AoI grows above a certain threshold that is a decreasing function of $q$. On the other hand, if $q>\frac{1}{2}$, the optimal policy turns {\it greedy}, in which a new update is transmitted whenever energy is available. Intuitively, the higher the erasure probability, the more eager the sensor becomes to transmit new updates, so that when they are eventually received successfully the AoI is not too large.

\section{System Model and Problem Formulation}

Consider an energy harvesting sensor that is sending status updates regarding measurements of some physical phenomenon to a destination. Updates are prone to erasures with some probability $q\in(0,1)$, representing the case in which the destination is unable to decode the status update message, due to fading or noise contamination in the channel. Such erasure events occur independently of the transmitted status updates and are also mutually independent over time. However, when a status update does go through with no erasures, it arrives at the destination instantly as in \cite{jing-age-online, arafa-age-online-finite, elif-age-online-threshold}. Each status update carries a time stamp denoting when it was acquired at the sensor. From the destination's perspective, the AoI at time $t$, $a(t)$, is defined as the time elapsed since the latest update has been {\it successfully} received, i.e., with no erasures. This is mathematically given by
\begin{align}
a(t)=t-u(t),
\end{align}
where $u(t)$ is the time stamp of the last update that has been successfully received prior to time $t$.

Energy expenditure is normalized in the sense that each status update transmission (and measurement/processing) consumes $1$ energy unit. In this work, we investigate the case in which the sensor is equipped with a {\it unit-sized battery} to save its incoming energy. This special case models situations in which a single transmission (along with its processing requirements) from the sensor completely depletes its battery. Energy arrives (is harvested) according to a Poisson process with (normalized) rate $1$. The setting is {\it online}, in which arrival times are revealed causally after the energy is harvested.

Let $l_i$ denote the time at which the sensor measures and transmits its $i$th update, and denote by $\mathcal{E}(t)$ the energy available in the battery at time $t$. Then, we must have the following energy causality constraint:
\begin{align} \label{eq_en_caus}
\mathcal{E}\left(l_i^-\right)\geq1,\quad\forall i.
\end{align}
Under any policy satisfying the above, the battery state evolves as follows:
\begin{align} \label{eq_btry_ev}
\mathcal{E}\left(l_i^-\right)=\min\{\mathcal{E}\left(l_{i-1}^-\right)-1+A_i,1\},\quad\forall i,
\end{align}
where $A_i$ denotes the amount of energy harvested in the time interval $[l_{i-1},l_i)$, which has a Poisson distribution with parameter $l_i-l_{i-1}$. Without loss of generality, we assume $l_0=0$, $\mathcal{E}(0)=1$, and that $a(0)=0$. That is, the system starts at time $0$ with {\it fresh} information.

\begin{figure}[t]
\center
\includegraphics[scale=1]{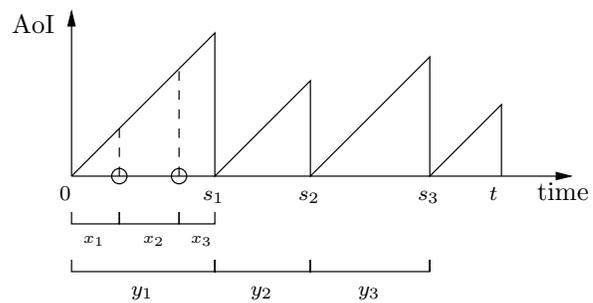}
\caption{Age evolution versus time with $n(t)=3$ successful updates. Circles denote failed attempts. In this example, the first update is successfully received after three update attempts.}
\label{fig_age_xmpl_erasure}
\end{figure}

In this work, we focus on the case in which there is {\it no erasure feedback} sent to the sensor from the destination. In other words, the sensor is oblivious to what occurs to its transmitted updates in the channel. However, the sensor is aware of the fact that updates are subject to erasures with probability $q$, and takes that probability knowledge into consideration while deciding on when to transmit. Let $s_i$ denote the $i$th {\it actual} update time at the destination, i.e., the time at which the $i$th status update has successfully reached the destination. Then, we have $s_0=l_0=0$ by assumption, and, in general, $\{s_i\}\subseteq\{l_i\}$. Let $x_i\triangleq l_i-l_{i-1}$ and $y_i\triangleq s_i-s_{i-1}$ denote the inter-update attempt and the actual inter-update delays, respectively, and let $n(t)$ denote the number of updates that are successfully received by time $t$. We are interested in the average AoI given by the area under the age curve, see Fig.~\ref{fig_age_xmpl_erasure}, which is given by
\begin{align} \label{eq_aoi_area}
r(t)=\frac{1}{2}\sum_{i=1}^{n(t)}y_i^2+\frac{1}{2}\left(t-s_{n(t)}\right)^2.
\end{align}

The goal is to devise online feasible status update policies that minimize the long term average AoI through choosing the transmission times $l_i$'s (or equivalently the $x_i$'s). That is, to solve the following problem:
\begin{align} \label{opt_no_fb}
\min_{\{x_i\}}\quad&\limsup_{T\rightarrow\infty}\frac{1}{T}\mathbb{E}\left[r(T)\right] \nonumber \\
\mbox{s.t.}\quad &(\ref{eq_en_caus})-(\ref{eq_btry_ev}),
\end{align}
under causal knowledge of the energy arrival times, with no erasure feedback, and given the erasure probability $q$.

\section{Key Characteristic of the Optimal Solution:\\Renewal Structure}

In this section, we discuss a key characteristic of the optimal solution of problem (\ref{opt_no_fb}). Specifically, we show that the optimal status update policy is a renewal policy, in which the {\it actual} inter-update times $y_i$'s are i.i.d. and that the {\it actual} update times $s_i$'s form a renewal process.

\begin{figure}
\center
\includegraphics[scale=.85]{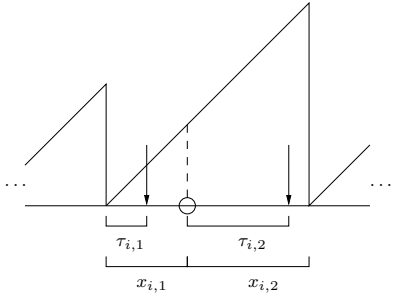}
\caption{AoI in the $i$th epoch with two update attempts. Arrows represent energy arrivals, and the circle denotes a failed update attempt.}
\label{fig_no_fb_age_epoch_tau}
\end{figure}

We first define some terminologies and notations. We use the term {\it epoch} to denote the time in between two consecutive successful updates. For instance, the $i$th epoch starts at time $s_{i-1}$ and ends at $s_i$, and has a length of $y_i$ time units. Note that an epoch may contain more than one update attempt, and the number of update attempts may vary from one epoch to another. We now slightly change notation to fit into our epoch definition and denote by $x_{i,k}$ the time in between the $(k-1)$th and the $k$th update attempt in the $i$th epoch. Similarly, let $\tau_{i,k}$ denote the time until the $k$th energy arrival in the $i$th epoch {\it starting from the $(k-1)$th update attempt}. For example, the first energy arrival in the $i$th epoch occurs at $s_{i-1}+\tau_{i,1}$, after which an update attempt occurs at $s_{i-1}+x_{i,1}$, with $x_{i,1}\geq\tau_{i,1}$ due to energy causality. Now say that this first update attempt has failed. Then, the sensor waits for the second energy arrival in the epoch occurring at $s_{i-1}+x_{i,1}+\tau_{i,2}$, after which the second update attempt occurs at $s_{i-1}+x_{i,1}+x_{i,2}$, and so on. Note that according to the definition of $\tau_{i,k}$'s, they do not necessarily represent the energy inter-arrival times, since $x_{i,k}$ can be strictly larger than $\tau_{i,k}$ (see Fig.~\ref{fig_no_fb_age_epoch_tau}).

Observe that transmission attempts occurring in the $i$th epoch may depend, in principal, on the history of events (transmission attempts and energy arrivals) that had occurred before the epoch started, which we denote by $\mathcal{H}_{i-1}$. Theorem~\ref{thm_rnwl} below shows, under some regularity conditions, that this is not the case; events in an epoch are independent of the history of events in previous epochs. Before we make that statement precise, we focus on the following special case of online policies, which are also the focus in \cite{jing-age-online, arafa-age-online-finite}:
\begin{definition}[Uniformly Bounded Policy] \label{def_ubp}
An online policy whose inter-update times have a bounded second moment.
\end{definition}
Intuitively, one would expect practical status update policies to be uniformly bounded as per Definition~\ref{def_ubp}, so that the inter-update delays do not grow infinitely large (in expectation). We now state the main result of this section in Theorem~\ref{thm_rnwl} below. The proof of the theorem, which is fully presented in the Appendix, is similar in essence to the proofs of \cite[Theorems 1 and 2]{arafa-age-online-finite} albeit some notable differences.

\begin{theorem} \label{thm_rnwl}
In the optimal solution of problem (\ref{opt_no_fb}), any uniformly bounded policy is outperformed by a renewal policy in which the epoch lengths, $y_i$'s, are i.i.d.
\end{theorem}

\section{Optimal Renewal Policies:\\Threshold Policies}

We now analyze the best renewal policy and show that it has a threshold structure. Theorem~\ref{thm_rnwl} shows that epoch starting times, $s_i$'s, at which the system resets by making both the sensor's battery and the AoI drop to $0$ simultaneously, constitute a renewal process. Since epoch lengths are i.i.d., we drop the subscript $i$ from all random variables and denote the epoch duration by $y$ and the inter-update attempt duration by $x$. Observe that we do not differentiate between different update attempts in a single epoch since the sensor is unaware of this information due to lack of erasure feedback. From the sensor's point of view, it only designs a single inter-update attempt duration $x$. However, it takes the value of $q$ into account while doing so as we show in the sequel. Let us (re)define $\tau$ as the time elapsed until the next energy arrival starting from the previous update attempt. Given that the sensor is unaware of erasure events, the sensor is ignorant of when the epoch ends. Therefore, inter-update attempt times in the epoch are functions of only the most recent energy arrival time. That is, $x$ is only a function of $\tau$.

By the strong law of large numbers for renewal processes (the renewal reward theorem) \cite{ross_stochastic}, problem (\ref{opt_no_fb}) now reduces to an optimization over a single epoch as follows:
\begin{align} \label{opt_no_fb_rnwl}
\min_{x(\cdot)}\quad&\frac{\mathbb{E}\left[R\right]}{\mathbb{E}\left[y\right]} \nonumber \\
\mbox{s.t.}\quad&x(\tau)\geq\tau,\quad\forall\tau,
\end{align}
where the random variable $R$ denotes the area under the age curve in the epoch. We now introduce the following auxiliary problem to solve the above:
\begin{align} \label{opt_no_fb_aux}
p(\lambda)\triangleq\min_{x(\cdot)}\quad&\mathbb{E}\left[R\right]-\lambda\mathbb{E}\left[y\right] \nonumber \\
\mbox{s.t.}\quad&x(\tau)\geq\tau,\quad\forall\tau,
\end{align}
for some $\lambda\geq0$. One can show that the optimal solution of problem (\ref{opt_no_fb_rnwl}) is given by $\lambda^*$ that solves $p(\lambda^*)=0$ and that such $\lambda^*$ is unique since $p(\lambda)$ is decreasing in $\lambda$ \cite{dinkelbach-fractional-prog}. Before solving problem (\ref{opt_no_fb_aux}), we first evaluate the terms $\mathbb{E}\left[y\right]$ and $\mathbb{E}\left[R\right]$ in what follows.

The expected epoch length can be found using iterated expectations by conditioning on how many erasure events occurred in it. We now write the following:
\begin{align}
\mathbb{E}\left[y\right]=&(1-q)\mathbb{E}\left[x(\tau_1)\right] \nonumber \\
&+q(1-q)\left(\mathbb{E}\left[x(\tau_1)\right]+\mathbb{E}\left[x(\tau_2)\right]\right) \nonumber \\
&+q^2(1-q)\left(\mathbb{E}\left[x(\tau_1)\right]+\mathbb{E}\left[x(\tau_2)\right]+\mathbb{E}\left[x(\tau_3)\right]\right) \nonumber \\
&+\dots \\
=&\mathbb{E}\left[x(\tau)\right]\left(1+q+q^2+\dots\right) \\
=&\frac{\mathbb{E}\left[x(\tau)\right]}{1-q}, \label{eq_no_fb_L}
\end{align}
where $\tau\sim\text{exp}(1)$, and the second equality follows since $\tau_j$'s are i.i.d. $\text{exp}(1)$ random variables by the memoryless property of the exponential distribution. The expected area under the age curve in a single epoch can be found similarly as follows:
\begin{align}
\mathbb{E}\left[R\right]=&(1-q)\frac{1}{2}\mathbb{E}\left[x^2(\tau_1)\right] \nonumber \\
&+q(1-q)\frac{1}{2}\mathbb{E}\left[\left(x(\tau_1)+x(\tau_2)\right)^2\right] \nonumber \\
&+q^2(1-q)\frac{1}{2}\mathbb{E}\left[\left(x(\tau_1)+x(\tau_2)+x(\tau_3)\right)^2\right] \nonumber \\
&+\dots \\
=&\frac{1}{2}\mathbb{E}\left[x^2(\tau)\right]\left(1+q+q^2+\dots\right) \nonumber \\
&+\left(\mathbb{E}\left[x(\tau)\right]\right)^2\left(q+2q^2+3q^3+\dots\right) \\
=&\frac{\frac{1}{2}\mathbb{E}\left[x^2(\tau)\right]}{1-q}+\frac{q\left(\mathbb{E}\left[x(\tau)\right]\right)^2}{(1-q)^2}, \label{eq_no_fb_R}
\end{align}
where the second equality again follows since $\tau_j$'s are i.i.d., and after some algebraic manipulations.

Using (\ref{eq_no_fb_L}) and (\ref{eq_no_fb_R}), one can write the following Lagrangian \cite{boyd} for problem (\ref{opt_no_fb_aux}):
\begin{align}
\mathcal{L}=&\frac{\frac{1}{2}\mathbb{E}\left[x^2(\tau)\right]}{1-q}+\frac{q\left(\mathbb{E}\left[x(\tau)\right]\right)^2}{(1-q)^2}-\lambda\frac{\mathbb{E}\left[x(\tau)\right]}{1-q} 
\nonumber \\
&-\int_0^\infty \left(x(\tau)-\tau\right)\eta(\tau)d\tau,
\end{align}
where $\eta$ is a Lagrange multiplier. Taking (the functional) derivative with respect to $x(t)$ and equating to $0$, we get that the optimal $x$ satisfies
\begin{align}
x(t)=\lambda-\frac{2q}{1-q}\mathbb{E}\left[x(\tau)\right]+\frac{\eta(t)}{e^{-t}/1-q}.
\end{align}
Now let us define
\begin{align} \label{eq_no_fb_lmda_prm_ini}
\lambda^\prime \triangleq \lambda-\frac{2q}{1-q}\mathbb{E}\left[x(\tau)\right].
\end{align}
The sign of $\lambda^\prime$ has a major implication on the optimal policy's structure, which we discuss in detail next.

If $\lambda^\prime<0$ then we must have $\eta(t)>0$, $\forall t$, to maintain positivity of $x(t)$. By complementary slackness \cite{boyd} this further implies that $x(t)=t$, $\forall t$, i.e., a {\it greedy zero-wait policy} is optimal in this case, in which energy is used to send an update whenever it arrives. This case occurs for relatively high values of $q$ which we specify precisely towards the end of this section. The value of $p(\lambda)$ in this case can be computed by plugging in $x(\tau)=\tau$ with $\mathbb{E}\left[x(\tau)\right]=1$ and $\mathbb{E}\left[x^2(\tau)\right]=2$ to get after some direct manipulations that
\begin{align}
p(\lambda)=\frac{1-\lambda(1-q)}{(1-q)^2},
\end{align}
which admits an optimal long term average AoI, $\lambda^*$, of
\begin{align} \label{eq_greedy_aoi}
\lambda^*=\frac{1}{1-q}.
\end{align}
Note that such greedy policy is always feasible and therefore (\ref{eq_greedy_aoi}) can generally serve as an upper bound on $\lambda^*$.

Now if $\lambda^\prime\geq0$, then by complementary slackness \cite{boyd} we get that (see \cite{jing-age-online} and \cite{arafa-age-online-finite})
\begin{align} \label{eq_no_fb_x}
x(t)=\begin{cases}\lambda^\prime,\quad &t<\lambda^\prime\\ t,\quad &t\geq\lambda^\prime \end{cases}.
\end{align}
That is, the optimal status update policy is a $\lambda^\prime$-{\it threshold policy}. Using this, one can directly compute $\mathbb{E}\left[x(\tau)\right]=\lambda^\prime+e^{-\lambda^\prime}$ and substitute back in (\ref{eq_no_fb_lmda_prm_ini}) to get that
\begin{align} \label{eq_no_fb_lmda_prm}
\frac{1+q}{1-q}\lambda^\prime+\frac{2q}{1-q}e^{-\lambda^\prime}=\lambda.
\end{align}
Direct first derivative analysis shows that the left hand side above is increasing in $\lambda^\prime$ for $\lambda^\prime\geq0$, and therefore, since its value at $\lambda^\prime=0$ is $2q/(1-q)$, (\ref{eq_no_fb_lmda_prm}) has a unique solution in $\lambda^\prime$ for every given $\lambda\geq2q/(1-q)$, i.e., $2q/(1-q)$ is the best achievable long term average AoI if $\lambda^\prime\geq0$. Now observe that for $q>1/2$, the greedy zero-wait policy achieves a lower long term average AoI than that, given by $1/(1-q)$. We therefore conclude that in the optimal policy, $\lambda^\prime$ can only be non-negative if $q\leq1/2$. Continuing with this assumption, we use (\ref{eq_no_fb_x}), and some algebraic manipulations, to get
\begin{align} \label{eq_no_fb_p}
p(\lambda^\prime)=\frac{(1-q)\left(e^{-\lambda^\prime}-\frac{1}{2}\left(\lambda^\prime\right)^2\right)-q\left(\lambda^\prime+e^{-\lambda^\prime}\right)^2}{(1-q)^2},
\end{align}
with $\lambda^\prime$ as defined in (\ref{eq_no_fb_lmda_prm}). Now observe that solving $p\left(\lambda^\prime\right)=0$ for $\lambda^\prime\geq0$ is tantamount to having $p(0)\geq0$ (since $p(\lambda)$ is monotonically decreasing \cite{dinkelbach-fractional-prog} in $\lambda$, and $\lambda$ is an increasing function of $\lambda^\prime$ from (\ref{eq_no_fb_lmda_prm})). In other words, we must have
\begin{align}
p(0)=\frac{1-2q}{(1-q)^2}\geq0 \iff q\leq\frac{1}{2}
\end{align}
as assumed before.

In conclusion, the optimal policy's structure depends on the value of the erasure probability, $q$. If $q>\frac{1}{2}$ then (\ref{eq_no_fb_p}) does not admit a positive $\lambda^\prime$ solution for $p\left(\lambda^\prime\right)=0$, and therefore it holds that $\lambda^\prime<0$, and the greedy zero-wait policy is optimal. While if $q\leq\frac{1}{2}$ then the optimal policy is a $\lambda^\prime$-threshold policy as in (\ref{eq_no_fb_x}), with the optimal $\lambda^\prime$ solving $p\left(\lambda^\prime\right)=0$. 

We conclude this section by stating a few remarks and presenting some numerical examples to further illustrate the results of this paper. First, observe that for the case of no erasures, i.e., $q=0$, (originally considered in \cite{jing-age-online}) we get from (\ref{eq_no_fb_lmda_prm}) and (\ref{eq_no_fb_p}) that $\lambda^\prime=\lambda$ and $p(\lambda)=e^{-\lambda}-\frac{1}{2}\lambda^2$, respectively, coinciding with the optimal solution in \cite{jing-age-online}. Second, for a given $\lambda\geq0$, (\ref{eq_no_fb_lmda_prm}) shows that $\lambda^\prime\leq\lambda$ with equality if and only if $q=0$. This shows that the problem with erasures does {\it not} have the recurring property shown in \cite{jing-age-online, arafa-age-online-finite, elif-age-online-threshold} that the optimal long term average AoI equals the optimal threshold; they are only equal if $q=0$.

\begin{figure}[t]
\center
\includegraphics[scale=.45]{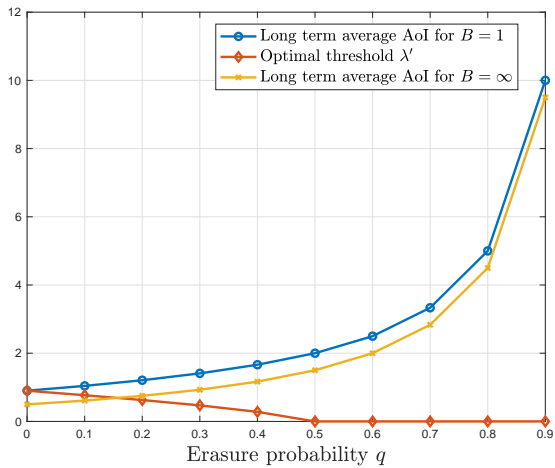}
\caption{Comparison of optimal AoI for $B=1$ and $B=\infty$, along with $\lambda^\prime$, versus $q$.}
\label{fig_no_fb_age_q}
\end{figure}

As for the numerical examples, in Fig.~\ref{fig_no_fb_age_q} we plot the optimal AoI versus the erasure probability, along with the corresponding optimal threshold. For the case $q\leq\frac{1}{2}$, we basically start with a large enough value of $\lambda^\prime$ that makes $p\left(\lambda^\prime\right)<0$, and then use a bisection search (in between $0$ and that large enough value) to find $\lambda^\prime$ that solves $p\left(\lambda^\prime\right)=0$. We then use (\ref{eq_no_fb_lmda_prm}) to find the optimal long term average AoI, $\lambda^*$. We also plot the optimal long term average AoI for the case $B=\infty$, which is shown in \cite{jing-age-error-infinite-no-fb} to be equal to $\frac{1+q}{2(1-q)}$. Clearly, the solution for the $B=\infty$ case serves as a lower bound for the solution for the $B=1$ case. From the figure, we also see that, quite intuitively, the larger the erasure probability, the larger the AoI, i.e., $\lambda^*$ is monotonically increasing in $q$. In addition, we see that the optimal threshold $\lambda^\prime$ is monotonically decreasing in $q$. This is quite intuitive, since the sensor should be more eager to send new updates if the erasure probability is high, so that when the update is eventually received successfully the AoI would not be large.

\section{Conclusion}

The long term average age-minimal online policy has been derived for an energy harvesting sensor with unit-sized battery, communicating with a destination over an erasure channel, without erasure error feedback information. Firstly, a key structure of the optimal policy has been shown, namely, that transmission times should be such that updates reach the destination successfully at times that constitute a renewal process. Then, it has been shown that the optimal renewal-type policy has an erasure probability-dependent threshold structure, in which a new update is transmitted only if the AoI grows above a certain threshold value. The optimal threshold has been shown to be decreasing with the erasure probability, reaching exactly $0$ when the erasure probability is $\frac{1}{2}$, and staying at $0$ afterwards. Such $0$-threshold policies are interpreted as greedy policies, in which a new update is sent whenever an energy arrives. The rationale is that with higher erasure probability the sensor should be more aggressive in transmitting new updates, so that when they eventually reach the destination successfully the AoI would not be too large.

\appendix[Proof of Theorem~\ref{thm_rnwl}] \label{apndx_thm_rnwl}

We follow an indirect approach to prove the theorem. Basically, we derive an achievable lower bound on the long term average AoI using renewal-type policies in a {\it genie-aided} system in which there exists a genie that informs the sensor when updates are successful, i.e., the epochs' start times. However, we enforce a constraint on the sensor {\it not to use} the lack of this piece of information to infer that its update is unsuccessful and act accordingly to change its policy within the same epoch. This seemingly unintuitive constraint simplifies the proof as we will see later on. Now observe that such genie-aided system cannot perform worse than the original system that we consider in this paper, and hence, a lower bound on this genie-aided system is also a lower bound on the original one. We then conclude the proof by showing that such lower bound is also achievable in the original system by showing that the optimal renewal-type policy does not actually need the information provided by the genie, thereby proving optimality of renewal-type policies in the original system as well. Next, we provide the details.

In the genie-aided system, consider any online feasible uniformly bounded policy $\{x_{i,k}\}$. Focusing on the $i$th epoch, let us denote by $R_{i,m}$ the area under the age curve in the $i$th epoch given that it went through $m$ update attempts, and by $R_i$ the area under the age curve in it {\it irrespective} of how many update attempts. Let us also denote by $e_{i,k}$ the event that the $k$th update attempt in the $i$th epoch gets erased. We can now write the following:
\begin{align}
R_{i,m}=&\frac{1}{2}\left(x_{i,1}+x_{i,2}+\dots+x_{i,m}\right)^2, \\
R_i=&\sum_{m=1}^\infty R_{i,m}\cdot\prod_{k=1}^{m-1}\mathbbm{1}\left(e_{i,k}\right)\mathbbm{1}\left(e_{i,m}^c\right), \label{eq_r_rm}
\end{align}
where $\mathbbm{1}(\cdot)$ is the indicator function, and the superscript $c$ denotes the complement of an event.

Next, for a fixed time $T$, denote by $N_T$ the number of epochs that have already {\it started} by time $T$. Given the history before the $i$th epoch, $\mathcal{H}_{i-1}$, and the number of update attempts in the $i$th epoch, $m$, let us define the vector ${\bm \tau}_i^{(m)}\triangleq[\tau_{i,1},\tau_{i,2},\dots,\tau_{i,m}]$, and define the following statistical average of the area under the age curve in the $i$th epoch with $m$ update attempts:
\begin{align}
\hat{R}_{i,m}\left({\bm \gamma}^{(m)},\mathcal{H}_{i-1}\right)\triangleq\mathbb{E}\left[R_{i,m}\Big|{\bm \tau}_i^{(m)}={\bm \gamma}^{(m)},\mathcal{H}_{i-1}\right].
\end{align}
Therefore, it holds that
\begin{align} \label{eq_rm_itr_exp}
&\hspace{-.075in}\mathbb{E}\left[R_{i,m}\mathbbm{1}\left(i\leq N_T\right)\right] \nonumber \\
&\hspace{-.075in}=\mathbb{E}_{\mathcal{H}_{i-1}}\!\!\left[\mathbb{E}_{{\bm \tau}_i^{(m)}}\!\!\left[\hat{R}_{i,m}\!\left(\!{\bm \gamma}^{(m)},\mathcal{H}_{i-1}\!\right)\right]\!\mathbbm{1}\left(i\leq N_T\right)\Big|\mathcal{H}_{i-1}\right]
\end{align}
since $\mathbbm{1}\left(i\leq N_T\right)$ is independent of ${\bm \tau}_i^{(m)}$ given $\mathcal{H}_{i-1}$. We can similarly define the following statistical average of the $i$th epoch length with $m$ update attempts:
\begin{align}
\hat{y}_{i,m}\left({\bm \gamma}^{(m)},\mathcal{H}_{i-1}\right)\triangleq\mathbb{E}\left[y_{i,m}\Big|{\bm \tau}_i^{(m)}={\bm \gamma}^{(m)},\mathcal{H}_{i-1}\right].
\end{align}

Next, observe that by (\ref{eq_aoi_area}) the following holds:
\begin{align} \label{eq_aoi_area_bd}
\hspace{-.1in}\frac{1}{T}\sum_{i=1}^\infty\! R_i\mathbbm{1}\left(i\leq N_T-1\right)\!\leq\!\frac{r(T)}{T}\!\leq\!\frac{1}{T}\sum_{i=1}^\infty\! R_i\mathbbm{1}\left(i\leq N_T\right).
\end{align}
\begin{figure*}
\begin{align}
\frac{1}{T}\mathbb{E}&\left[\sum_{i=1}^\infty R_i\mathbbm{1}\left(i\leq N_T\right)\right] \nonumber \\
\geq&\frac{\mathbb{E}\left[\sum_{i=1}^\infty R_i\mathbbm{1}\left(i\leq N_T\right)\right]}{\mathbb{E}\left[\sum_{i=1}^\infty y_i\mathbbm{1}\left(i\leq N_T\right)\right]} \label{eq_pf_1} \\
=&\frac{\mathbb{E}\left[\sum_{i=1}^\infty \sum_{m=1}^\infty R_{i,m}\prod_{k=1}^{m-1}\mathbbm{1}\left(e_{i,k}\right)\mathbbm{1}\left(e_{i,m}^c\right)\mathbbm{1}\left(i\leq N_T\right)\right]}{\mathbb{E}\left[\sum_{i=1}^\infty y_i\mathbbm{1}\left(i\leq N_T\right)\right]} \label{eq_pf_2} \\
=&\frac{\sum_{i=1}^\infty \sum_{m=1}^\infty q^{m-1}(1-q)\mathbb{E}\left[R_{i,m}\mathbbm{1}\left(i\leq N_T\right)\right]}{\sum_{i=1}^\infty\mathbb{E}\left[y_i\mathbbm{1}\left(i\leq N_T\right)\right]} \label{eq_pf_3} \\
=&\frac{\sum_{i=1}^\infty \sum_{m=1}^\infty q^{m-1}(1-q) \mathbb{E}_{\mathcal{H}_{i-1}}\left[\mathbb{E}_{{\bm \tau}_i^{(m)}}\left[\hat{R}_{i,m}\left({\bm \gamma}^{(m)},\mathcal{H}_{i-1}\right)\right]\mathbbm{1}\left(i\leq N_T\right)\Big|\mathcal{H}_{i-1}\right]}{\sum_{i=1}^\infty\mathbb{E}\left[y_i\mathbbm{1}\left(i\leq N_T\right)\right]} \label{eq_pf_4} \\
=&\frac{\sum_{i=1}^\infty \mathbb{E}_{\mathcal{H}_{i-1}}\left[\sum_{m=1}^\infty q^{m-1}(1-q)\mathbb{E}_{{\bm \tau}_i^{(m)}}\left[\hat{R}_{i,m}\left({\bm \gamma}^{(m)},\mathcal{H}_{i-1}\right)\right]\mathbbm{1}\left(i\leq N_T\right)\Big|\mathcal{H}_{i-1}\right]}{\sum_{i=1}^\infty\mathbb{E}\left[y_i\mathbbm{1}\left(i\leq N_T\right)\right]} \label{eq_pf_5} \\
=&\frac{\sum_{i=1}^\infty \mathbb{E}_{\mathcal{H}_{i-1}}\left[ \sum_{m=1}^\infty q^{m-1}(1-q)\mathbb{E}_{{\bm \tau}_i^{(m)}}\left[\hat{y}_{i,m}\left({\bm \gamma}^{(m)},\mathcal{H}_{i-1}\right)\right] \frac{\sum_{m=1}^\infty q^{m-1}(1-q)\mathbb{E}_{{\bm \tau}_i^{(m)}}\left[\hat{R}_{i,m}\left({\bm \gamma}^{(m)},\mathcal{H}_{i-1}\right)\right]}{\sum_{m=1}^\infty q^{m-1}(1-q)\mathbb{E}_{{\bm \tau}_i^{(m)}}\left[\hat{y}_{i,m}\left({\bm \gamma}^{(m)},\mathcal{H}_{i-1}\right)\right]} \mathbbm{1}\left(i\leq N_T\right) \Big| \mathcal{H}_{i-1}\right]}{\sum_{i=1}^\infty\mathbb{E}\left[y_i\mathbbm{1}\left(i\leq N_T\right)\right]} \label{eq_pf_6} \\
\geq&\frac{\sum_{i=1}^\infty \mathbb{E}_{\mathcal{H}_{i-1}}\left[ \sum_{m=1}^\infty q^{m-1}(1-q)\mathbb{E}_{{\bm \tau}_i^{(m)}}\left[\hat{y}_{i,m}\left({\bm \gamma}^{(m)},\mathcal{H}_{i-1}\right)\right] R^*\left(\mathcal{H}_{i-1}\right) \mathbbm{1}\left(i\leq N_T\right) \Big| \mathcal{H}_{i-1}\right]}{\sum_{i=1}^\infty\mathbb{E}\left[y_i\mathbbm{1}\left(i\leq N_T\right)\right]} \label{eq_pf_7} \\
\geq&\frac{\sum_{i=1}^\infty \sum_{m=1}^\infty q^{m-1}(1-q) \mathbb{E}_{\mathcal{H}_{i-1}}\left[\mathbb{E}_{{\bm \tau}_i^{(m)}}\left[\hat{y}_{i,m}\left({\bm \gamma}^{(m)},\mathcal{H}_{i-1}\right)\right] \mathbbm{1}\left(i\leq N_T\right) \Big| \mathcal{H}_{i-1}\right]}{\sum_{i=1}^\infty\mathbb{E}\left[y_i\mathbbm{1}\left(i\leq N_T\right)\right]} R_{\min} \label{eq_pf_8} \\
=&\frac{\sum_{i=1}^\infty \sum_{m=1}^\infty q^{m-1}(1-q)\mathbb{E}\left[y_{i,m}\mathbbm{1}\left(i\leq N_T\right)\right]}{\sum_{i=1}^\infty\mathbb{E}\left[y_i\mathbbm{1}\left(i\leq N_T\right)\right]} R_{\min} \label{eq_pf_9} \\
=&R_{\min}. \label{eq_pf_10}
\end{align}
\hrulefill
\end{figure*}
\hspace{-.075in}Following similar analysis as in \cite[Appendix C-1]{jing-age-online}, one can show that the term $\mathbb{E}\left[R_{N_T}\right]/T\rightarrow0$ as $T\rightarrow\infty$, making the upper and lower bounds in (\ref{eq_aoi_area_bd}) equal as $T\rightarrow\infty$. Therefore, we proceed by deriving a lower bound on $\frac{1}{T}\mathbb{E}\left[\sum_{i=1}^\infty R_i\mathbbm{1}\left(i\leq N_T\right)\right]$ and conclude that it shall also serve as a lower bound on $\frac{1}{T}\mathbb{E}\left[r(T)\right]$ as $T\rightarrow\infty$ (the objective function of problem (\ref{opt_no_fb})). We do so through a series of inequalities at the top of the next page. There, (\ref{eq_pf_1}) follows since, by definition of $N_T$, it holds that $\mathbb{E}\left[\sum_{i=1}^\infty y_i\mathbbm{1}\left(i\leq N_T\right)\right]\geq T$; (\ref{eq_pf_2}) follows by (\ref{eq_r_rm}); (\ref{eq_pf_3}) follows by the monotone convergence theorem, and the fact that erasure events are mutually independent and are independent of transmissions; (\ref{eq_pf_4}) follows by (\ref{eq_rm_itr_exp}); (\ref{eq_pf_5}) follows again by the monotone convergence theorem; $R^*\left(\mathcal{H}_{i-1}\right)$ in (\ref{eq_pf_7}) denotes the minimum value of $\frac{\sum_{m=1}^\infty q^{m-1}(1-q)\mathbb{E}_{{\bm \tau}_i^{(m)}}\left[\hat{R}_{i,m}\left({\bm \gamma}^{(m)},\mathcal{H}_{i-1}\right)\right]}{\sum_{m=1}^\infty q^{m-1}(1-q)\mathbb{E}_{{\bm \tau}_i^{(m)}}\left[\hat{y}_{i,m}\left({\bm \gamma}^{(m)},\mathcal{H}_{i-1}\right)\right]}$; $R_{\min}$ in (\ref{eq_pf_8}) denotes the minimum value of $R^*\left(\mathcal{H}_{i-1}\right)$ over all epochs and their corresponding histories, i.e., the minimum over all $i$ and $\mathcal{H}_{i-1}$; and (\ref{eq_pf_9}) and (\ref{eq_pf_10}) follow by the relationships between $\hat{y}_{i,m}$, $y_{i,m}$, and $y_i$ which are the same as those between $\hat{R}_{i,m}$, $R_{i,m}$, and $R_i$ that got us from (\ref{eq_pf_1}) to (\ref{eq_pf_4}).

Note that the online policy achieving $R^*\left(\mathcal{H}_{i-1}\right)$ is only a function of the energy arrivals in the $i$th epoch, since the history $\mathcal{H}_{i-1}$ is fixed. Now observe that by the memoryless property of exponential distribution, $\tau_{i,k}$'s are i.i.d.$\sim\exp(1)$. Therefore, if one repeats the policy that achieves $R_{\min}$ over all epochs, which is possible since the genie provides information about epochs' start times, then one gets a renewal policy in which $y_i$'s are i.i.d.

We now argue that the best renewal policy does not depend on the genie's provided information. First, it is clear that when an epoch starts, the sensor's next inter-update attempt becomes independent of the past and only a function of the energy arrivals in the epoch, in particular the first arrival time. If the sensor receives an information from the genie that its first update was successful, then this means a new epoch started and the process is repeated. On the other hand, if it does not hear from the genie, it is not allowed to act upon that information according to our enforced constraint that we stated at the beginning of the proof. Hence, it repeats the same policy, otherwise the constraint would be violated. Therefore, the policy does not change whether the genie sends its information or not. Finally, observe that this policy is achievable in the original system considered in this paper, i.e., the system with no genie. This completes the proof.


\end{document}